\documentstyle[aps]{revtex}
\def\pint{\hbox{\ooalign{$\displaystyle\int$\cr$-$}}}
\begin{document}
\draft
\author{Robert de Mello Koch and Jo\~ao P. Rodrigues
}
\address{Centre for Nonlinear Studies and Department of Physics,\newline
University of the Witwatersrand,PO Wits 2050, Johannesburg,
South Africa}
\title{
\hspace{3in}
{\sc CNLS 94-03} {\tt (hep-th/9410012)} \vspace{0.2in}  \newline
The Collective Field theory of a Singular Supersymmetric
Matrix
Model}
\maketitle
\begin{minipage}{\textwidth}
\begin{quotation}
\begin{abstract}
The supersymmetric collective field theory with
the potential $v'(x)=\omega
x-{\eta\over x}$ is studied, motivated by the matrix
model proposed by Jevicki
and Yoneya to describe two dimensional string
theory in a black hole
background. Consistency with supersymmetry enforces
a two band solution. A
supersymmetric classical configuration is found, and
interpreted in terms  of the density of zeros of
certain Laguerre polynomials. The spectrum of the
model is then studied  and
is seen to correspond to a massless scalar
and a majorana fermion. The $x$
space eigenfunctions are constructed
and expressed in terms of Chebyshev
polynomials. Higher order interactions are also discussed.
\bigskip

\bigskip
\end{abstract}
\end{quotation}
\end{minipage}

\section{Introduction}
The collective field theory~\cite{JS} description of
the quantum mechanics of single matrix systems has had
important applications in QCD type models~\cite{JSa}
and in low dimensional strings~\cite{DaJ}. In particular
in the context of $c=1$ strings, the cubic collective
hamiltonian has been shown to reproduce perturbatively
the correlation functions of the theory~\cite{DJR} and
its exact fermionization~\cite{BIPZ} has been established
in the potential free case~\cite{AJ}.

We recall that the $c=1$ string interpretation is obtained
from the quantum mechanics of an hermitean matrix with
(inverted harmonic oscillator) potential

\begin{equation}
v_b = \mu - {1\over 2} x^2    \label{BosPot}
\end{equation}

in the double scaling limit~\cite{Gro,GrMi} $N \to \infty$ and
$\mu \to 0$ with $N\mu$ constant.

At the classical level, it is known that the collective field is
the density variable description of the dynamics of the $N$
eigenvalues of the matrix, which is described by a Calogero
type~\cite{OlPe} hamiltonian. This interplay between the collective
field description of matrix quantum mechanics and Calogero models
has played a fundamental role in the understanding of single matrix
theories: for instance it has enabled the identification of their
large $N$ limit with the classical equations of motion of the theory
subjected to specific initial conditions~\cite{JeLe}, and the fact
that the Calogero model with harmonic type potentials is exactly
integrable~\cite{OlPe} is at the root of the uncovering of the
$W_{\infty}$ algebra in the $c=1$ string~\cite{Pol,AvJe}.

A supersymmetric extension of the collective field theory has been
proposed by Jevicki and one of the authors based on the metric
structure of the cubic collective field theory ~\cite{JeRo}.
It is perhaps not surprising in the light of the discussion above
that
at the classical level this extension corresponds to a density
variable description of super-Calogero type models first discussed
by Friedman and Mende~\cite{FrMe} who specifically considered the
harmonic potential
\footnotetext[1]{Recall that the relationship between the
potential $v'(x)$ of the supersymmetric model and the bosonic
potential $v_b$ is $v_b={1\over 2}v'^{2}(x)+...$}

\begin{equation}
              v'(x) = \omega x   \label{HarPot}
\end{equation}
In a seemingly unrelated development, Marinari and
Parisi~\cite{MaPa} introduced a supersymmetric extension of single
matrix models by supertriangulation. It turns out, however, that the
restriction of this model to a suitable invariant subspace~\cite{Dab}
yields a system of the super-Calogero type. Therefore all three
approaches are essentially equivalent, as it has been established
in~\cite{RoVa}.

It was hoped that these approaches might provide a non-perturbative
formulation of the $\hat{c}=1$ superstring, but this has (so far)
turned out not to be possible. As a matter of fact, it was
established
{}~\cite{RoVa} that for a large class
of potentials one cannot generate an infinite Liouville-like
dimension as is the case of the $c=1$ model. This is ultimately
because it is hard to reconcile the positive definiteness of a
supersymmetric theory with the (essentially) negative definite
critical inverted harmonic oscillator potential of a $c=1$ theory
and the existence of a freely adjustable chemical potential parameter
$\mu$.
{}From a string theory point of view, these supersymmetric
extensions are perhaps more
relevant to the stabilization of $d=0$ models, in a well known
mechanism of dimensional reduction~\cite{GrHa} (for related
developments based on loop equations, please refer to~\cite{Al};
see also~\cite{Ram,Pol2} for recent discussions of other possible
continuum limits).

Recently, Jevicki and Yoneya~\cite{JeYo} suggested that a matrix
model
with potential

\begin{equation}
v_b = - {1\over 2} x^2 + {1\over 2} {M\over x^2} \label{BosBH}
\end{equation}

may be relevant to the study of two dimensional string theory
in a black-hole background, in the double scaling limit
$N\to \infty$ and $M\to 0$ with $N^2M$ constant. The existence of
a quantum mechanical model where black-hole physics
can be addressed in a possibly non-perturbative way cannot
be underestimated.
Moreover,
the matrix model with the potential~(\ref{BosBH}) has recently been
shown
to have an exact, computable S-matrix with particle
production~\cite{DKR} and is known to possess (with both $\pm x^2$
harmonic potential) an infinite Lie Algebra~\cite{JAAJ}. Also,
earlier studies of the physically acceptable
quantum mechanical solutions of the Schrodinger equation
for the $1\over x^2$ potential on the real line have been carried
out~\cite{Calo}. Therefore the study of
the deformation of the harmonic oscillator potential by
a singular $1\over x^2$ contribution is of great interest.

In this article we study the supersymmetric collective field
theory for the potential
\begin{equation}
v'(x) = \omega x - {\eta\over x}  \label{SusyBH}
\end{equation}

This corresponds to a supersymmetrization of the matrix model
with a bosonic potential which is, apart from a constant, given by

\begin{equation}
v_b = {1\over 2} \omega^2 x^2 + {1\over 2}
{\eta^2\over {x^2}}. \label{BBH}
\end{equation}

The potential~(\ref{SusyBH}) generalizes the potential~(\ref{HarPot})
originally  considered by Friedman and Mende~\cite{FrMe}. Via the the
supersymmetric collective field theory description, we obtain
a systematic semiclassical expansion  of the corresponding
super-Calogero
model which may potentially be of relevance to the study of
superstring
theory in a black-hole background.

After a brief review of supersymmetric collective
field theory~\cite{JeRo}
in Section 2, we carry out the analysis of the leading large-$N$
supersymmetry preserving configuration in Section 3. We show that
a symmetric density of eigenvalues exists which is
the minimum of the leading effective bosonic
potential and it simultaneously satisfies the usual analytic
properties of the corresponding $d=0$ solution to the problem.
The only other classical configuration known to
satisfy both requirements is that of the harmonic oscillator
potential~(\ref{HarPot}). This symmetric leading large $N$
configuration
turns out to have been
previously discussed by Tan~\cite{Tan} in a different context.
Using a classical result of Calogero~\cite{FCal}, we show that the
eigenvalues of the matrix correspond to the zeroes of an
appropriate Laguerre polynomial, generalizing a corresponding
result for the harmonic oscillator case and the zeroes of
Hermite polynomials~\cite{JeLe}.
In Section 4, we discuss small fluctuations about the leading
large $N$ configuration and show that the spectrum is linear.
We exhibit supersymmetry explicitly by identifying a
majorana fermion as the partner of the scalar field.
We also construct explicitly the polynomial eigenfunctions
which are seen to be related to the Chebyshev polynomials,
thereby generalizing a
class of polynomial solutions discussed by
Calogero~\cite{FC}. In Section 5, we discuss interactions and
reserve Section 6 for further discussion and conclusions.
We believe that the results presented in this article provide the
first collective field description
of a two band solution beyond tree level (multiband models
have been described at tree level and via orthogonal polynomials
in ref~\cite{Kres}).

\section{Collective field theory of a super symmetric matrix
 model}

 In~\cite{JeRo} the bosonic collective lagrangian:

\begin{equation}
L=\int dx\bigl( {1\over 2}{(\partial_{x}^{-1}\dot{\phi})^{2}
\over \phi}-{\pi^{2}\over 6}\phi^{3}(x,t)-v(x)\phi\bigr) \label{Bcl}
\end{equation}

 was supersymmetrized exploiting the metric structure of the
 kinetic energy term ($\phi\equiv\partial_{x}\varphi$)

\begin{eqnarray}
\nonumber
L_{T}&=&{1\over 2}\int dx {(\dot{\varphi})^{2}
\over \phi} \\
&=& \int dx \int dy \dot{\varphi} (x,t)g_{xy}(\varphi )
\dot{\varphi} (y,t) \label{Mst}
\end{eqnarray}

 The resulting supersymmetric lagrangian is

\begin{eqnarray}
\nonumber
L&=&{1\over 2}\int {dx\over\phi}\dot{\varphi}^{2}
-{1\over 2}\int dx \phi (W_{;x})^{2} \\
\nonumber
&+&{i\over 2}\int {dx \over \phi}(\psi^{\dagger}\dot{\psi}
-\dot{\psi^{\dagger}}\psi ) \\
\nonumber
&+&{i\over 2}\int {dx\over \phi}\dot{\varphi }
\bigl[\partial_{x}({\psi^{\dagger}\over\phi})\phi-\psi^\dagger
\partial_{x}({\psi \over\phi})\bigr] \\
\nonumber
&+&\int{dx\over\phi}\psi^{\dagger}\partial_{x}W_{;x}\psi\\
&-&\int dx \int dy \psi^{\dagger}(x)W_{;xy}\psi(y) \label{Susyl}
\end{eqnarray}

 where the superpotential $W(x)$ is given by

\begin{equation}
W(x)=\int dx v(x)\phi
-{1\over 2}\int dx\int dy
ln|x-y|\phi (x)\phi (y) \label{SP}
\end{equation}

 and $W_{;x}\equiv \partial_{x}\delta W[\phi ]/\delta\phi (x)
 =\delta W/\delta\varphi(x)$.
 This lagrangian is accompanied by the constraint

\begin{equation}
\int dx\phi(x) = N \label{Constraint}
\end{equation}

 The lagrangian~(\ref{Susyl}) is singular which is easily seen by
 computing
 the conjugate momenta of the fermionic fields. Due to the presence
 of
 the constraints

\begin{equation}
\chi = \Pi - {i\over 2}{\psi^{\dagger}\over\phi}, \quad
\bar{\chi} = \Pi^{\dagger}+{i\over 2}{\psi^{\dagger}\over\phi}
\label{Con}
\end{equation}

 quantization is achieved with the Dirac brackets

\begin{equation}
\{ A,B\}^{D}=\{ A,B\}-\sum_{i,j}\{ A,\chi_{i}\}(\{ \chi_{i},
\chi_{j}\}^{-1})_{ij}
\{\chi_{j},B\}. \label{Dbracket}
\end{equation}

Explicitly, after quantizing, one finds~\cite{JeRo}:

\begin{equation}
[p (x),\varphi (y)]=-i\delta (x-y) \label{Efcr}
\end{equation}

\begin{equation}
[p (x), \psi^{\dagger}(y)] = {i\over 2} {\psi^\dagger \over
\phi } (y)\partial_x \delta (x-y) \label{Escr}
\end{equation}

\begin{equation}
[p(x),\psi(y)] = {i\over 2} {\psi \over \phi }(y)\partial_x
\delta(x-y) \label{Etcr}
\end{equation}

\begin{equation}
\{\psi (x), \psi^\dagger (y) \} = \phi (x) \delta (x-y)
\label{Eocr}
\end{equation}

 The hamiltonian takes the simple form

\begin{eqnarray}
\nonumber
H&=&{1\over 2}\int dx\phi
(p(x)-{i\over 2}[\partial_{x}({\psi^{\dagger}\over\phi})
{\psi\over\phi}-{\psi^{\dagger}\over\phi}
\partial_{x}({\psi\over\phi})])^{2} \\
&&+{1\over 2}\int dx
\phi (W_{;x})^{2}\\
\nonumber
&&-\int {dx\over\phi}\psi^{\dagger}\partial_{x}W_{;x}\psi +
\int dx\int dy \psi^{\dagger}(x)W_{;xy}\psi(y)
\end{eqnarray}

 and the corresponding supercharges are

\begin{eqnarray}
\nonumber
Q &=& \int dx \psi^{\dagger}(x)[
(p(x)-\\
\nonumber
&&{i\over 2}
[\partial_{x}({\psi^{\dagger}\over\phi}){\psi\over\phi}
-{\psi^{\dagger}\over\phi}
\partial_{x}({\psi\over\phi})]
-iW_{;x}] \\
\nonumber
Q^{\dagger}&=&\int dx \psi (x)[
(p(x)-\\
&&{i\over 2}
[\partial_{x}({\psi^{\dagger}\over\phi}){\psi\over\phi}
-{\psi^{\dagger}\over\phi}
\partial_{x}({\psi\over\phi})]
+iW_{;x}]
\label{SupCharges}
\end{eqnarray}

 The above commutation relations~(\ref{Efcr}) -~(\ref{Eocr})
 are unusual in that the bosonic momentum does not commute
 with the fermionic fields. This is not however, hard to understand.
 Classically, the above fields correspond to a density description of
 a super Calogero model~\cite{JeLe,FrMe}:

\begin{equation}
\phi(x) = \sum_i \delta (x -
\lambda_i) \label{Efcf}
\end{equation}

\begin{equation}
\phi\sigma (x)= - \sum_i \delta (x- \lambda_i)p_i \label{Escf}
\end{equation}

\begin{equation}
\psi(x) = - \sum_i \delta (x - x_i)\psi_i \label{Etcf}
\end{equation}

\begin{equation}
\psi^\dagger (x)=-\sum_i\delta (x-x_i)\psi_i^\dagger. \label{Eocf}
\end{equation}

 where $\dot{\varphi}=\phi\sigma$.
 The $\lambda_{i}$'s are $N$ bosonic co-ordinates,
 $p_{i}$ are the corresponding momenta and $\psi_{i}$ and
 $\psi_{i}^{\dagger}$ are fermionic (Grassman) superpartners. It is
 easy to see that~(\ref{Efcr}) -~(\ref{Eocr}) also follows from
 these density fields.

 The commutators~(\ref{Efcr}) -~(\ref{Eocr})
 are brought into a more familiar form
 by rescaling~\cite{RoVa} $\psi (x)\to\sqrt{\phi (x)} \psi (x)$ and
 $\psi^\dagger (x)\to\sqrt{\phi (x)}\psi^\dagger (x)$.
 The new commutators are:

\begin{equation}
[\varphi (x), \varphi (y)]=0 \label{Enco}
\end{equation}

\begin{equation}
[\varphi (x), p(y)]=i \delta (x-y) \label{Enct}
\end{equation}

\begin{equation}
\{\psi (x), \psi^\dagger (y) \} = \delta (x-y) \label{Ench}
\end{equation}

\begin{eqnarray}
\nonumber
&[&\varphi (x),\psi (y)]=[\varphi (x),\psi^\dagger (y)]=[p(x),
\psi (y)]=\\
&=&[p(x), \psi^\dagger (y)]=0 \label{Encf}
\end{eqnarray}

 The hamiltonian becomes

\begin{eqnarray}
H &=& {1\over 2} \int {dx \over \phi} \Bigl( \phi p -
{i\over 2}\Bigl[
(\partial_x\psi^\dagger )\psi -
\psi^{\dagger}(\partial_{x}\psi )\Bigr] \Bigr)^2 \\
\nonumber
&&+{1\over 2}\int dx\phi
(W_{;x})^2 -{1\over 2} \int dx [
\psi^\dagger,\psi]\partial_x W_{;x}\\
\nonumber
&&{1\over 2}\int dx \int dy [\psi^\dagger (x),\psi (y)]
\sqrt{\phi (x)} W_{;xy} \sqrt{\phi (x)} \label{Erwh}
\end{eqnarray}

 and the supercharges:

\begin{eqnarray}
\nonumber
Q &=& \int dx \psi^\dagger (x) \sqrt {\phi (x)} \Bigl( p(x) -
  {i \over 2\phi}\Bigl( (\partial_x\psi^\dagger)\psi -
  \psi^\dagger (\partial_x \psi) \Bigr) \\
  &&+iv'(x)-i\pint dy
  {\partial_y \varphi\over x-y} \Bigr)\label{Ersq}
\end{eqnarray}

\begin{eqnarray}
\nonumber
Q^\dagger &=& \int dx \psi (x) \sqrt {\phi (x)} \Bigl( p(x) -
  {i \over 2\phi}\Bigl( (\partial_x\psi^\dagger)\psi -
  \psi^\dagger(\partial_x \psi) \Bigr) \\
  &&-iv'(x)+i\pint dy
  {\partial_y \varphi\over x-y} \Bigr) \label{Erss}
\end{eqnarray}

 The square root factors
 are treated by expanding about the large $N$ background
 configuration, which clearly generates an infinite
 perturbative expansion.

 Now, we turn our attention to the $N$ dependence of the
 Hamiltonian and the supercharges. It is well known that
 the Feynman diagrams of a matrix theory can be
 topologically classified according to their genus ~\cite{tHooft}.
 Writing the superpotential $\bar{v}$ in terms of the
 supermatrix $\Phi$~\cite{MaPa,RoVa}

\begin{equation}
\bar v(\Phi ) = Tr \Bigl( g_2\Phi^2+{g_3 \over \sqrt N}
\Phi^3+...+  {g_p \over N^{{p \over 2} -1}}\Phi^p\Bigr)\label{Elne}
\end{equation}

 it then follows that a diagram of genus $\Gamma$ carries a
 factor $N^{2-2\Gamma}$. It is clear that for $\bar v$ of the
 above form, the expressions
 ${v(\sqrt N x) \over \sqrt N}$ and $v'( \sqrt N x)$
 are both independent of $N$. Thus rescaling
 $x \to \sqrt N x$,$v \to {v \over\sqrt N}$
 and $\phi \to \sqrt N \phi$, our constraint on $\phi$ becomes
 $N$ independent ($\int dx \phi (x) = 1$) and the $N$ dependence
 of the Hamiltonian is made explicit:

\begin{eqnarray}
\nonumber
H &=& {N^2 \over 2}\int
dx \phi (x)\Bigl(\pint dy {\partial_y \varphi \over x-y}-v'(x)
\Bigr)^2 \\
\nonumber
&&+{1\over 2} \int dx \Bigl[ \psi^\dagger (x)
\sqrt{\phi (x)} ,{d \over dx} \pint dy {\psi (y)
\sqrt{\phi (y)} \over x-y}\Bigr] \label{Efgi}\\
\nonumber
&&-{1\over 2}\int dx [\psi^\dagger,\psi] {d \over dx}
\Bigl( \pint dy {\partial_y\varphi \over x-y} -v'(x) \Bigr) \\
&&+ {1 \over 2N^2} \int {dx \over \phi} \Bigl( \phi p -{i\over 2}
\Bigl[ (\partial_x\psi^\dagger ) \psi-\psi^{\dagger}
(\partial_{x}\psi )\Bigr)^2
\end{eqnarray}

 Requiring that supersymmetry is preserved to
 leading order, we obtain an integral equation for the
 vacuum configuration of the collective field:

\begin{equation}
\pint dy { \phi_0(y) \over x-y } - v'(x) = 0 \label{Epss}
\end{equation}

 Now, expand about the vacuum configuration $\phi_0$ as:

\begin{equation}
\phi = \phi_0 + {1 \over N} \partial_x \eta \label{Eevc}
\end{equation}

 and rescale $p \to Np$. The above rescaling of the bosonic
 momentum, together with the $1\over N$ coefficient of
 $\partial_x\eta$ ensures that all bosonic propagators are of
 order unity. Thus all explicit $N$ dependance is
 absorbed into the vertices. This rescaling is purely a
 matter of convenience and does not change the theory.

\section{The Vacuum Density}

 In this section
 we determine the (classical) vacuum for the
 superpotential~(\ref{SP})
 with

\begin{equation}
  v(x)={1\over 2}\omega x^2 -\eta N log|x|
\end{equation}

 The explicit factor of $N$ in the above equation is crucial for the
 identification of the large $N$ expansion with an expansion in the
 genus of the diagrams of the original supermatrix model (see the
 comments
 following~(\ref{Elne})). We rescale as explained in these comments,
 and are lead to the following integral equation for the vacuum
 configuration of the collective field

\begin{equation}
\pint dy{\phi(y)\over x-y }=v(x)=\omega x-{\eta \over x} \label{Ecfv}
\end{equation}

 with $\int dx \phi = 1$. We are interested in the solution
 for which $x$ runs from $-\infty$ to $+\infty$.
 This solution has been considered by Tan~\cite{Tan}
 in the context of the $d=0$ generalized Penner
 models and we repeat the derivation here. $v(x)$ has
 two minima, situated symmetrically about the origin. These
 minima are separated by an infinitly high potential barrier
 such that the tunneling probability from one region to the
 other is zero. We therefore demand that $\phi$ has support
 only over the two finite intervals, $y_-^2\le y^2\le y_+^2$,
 around the minima of $v(x)$ on the real $y$ axis.
 Thus, $\phi$ must not have a pole at the origin, since no
 eigenvalues can lie there. As is standard,
 we now continue into the
 complex plane, and denote the resulting function $G(z)$

\begin{equation}
G(z)=\pint dy{ \phi(y)\over z-y}
\end{equation}

 It follows that $G(z)$ only has two pairs of branch points at
 $y_{\pm}$ and $-y_{\pm}$.If we connect these pairs of branch
 points along the real axis, we obtain a real analytic function
 bounded everywhere on the first sheet. Finally we obtain a
 normalisation condition by considering the $|z|\to\infty$ limit,
 which upon using the constraint $\int dx \phi (x) = 1$ becomes
 $G(z)\sim {1 \over z}$. It is easy
 to show that a symmetric ansatz of the form :

\begin{equation}
G(z)= \omega z - {\eta \over z} + {d \over z}
\sqrt {(z^2-y_-^2)(z^2-y_+^2)} \label{Esvc}
\end{equation}

 leads to a consistent solution, with the normalization condition
 fixing $d=-\omega$ and

\begin{equation}
y_{\pm}^{2}={1+\eta\over\omega} \pm {1\over \omega}\sqrt{1+2\eta}
\end{equation}

 Now, observe that if we evaluate $G$ at
 $z^{\pm}=x{\pm} i\epsilon$ we obtain:

\begin{equation}
G(z^{\pm})=v(x){\pm}i\pi\phi (x) \label{Esvp}
\end{equation}

 from which it follows that the collective field is given by:

\begin{eqnarray}
\phi (x)&=& {1 \over |x|\pi}\sqrt{(1+2\eta)} \times \\
\nonumber
&&\sqrt{\Bigl( 1- {\omega^2 \over
 1+2\eta}(x^2-{1+\eta \over \omega})^2\Bigr)} \label{Esqh}
\end{eqnarray}

 We emphasize that in our case the density
 field $\phi (x)$ describes a double band of eigenvalues.
 Notice that this solution is an even function of $x$. The
 importance of the symmetry of the solution can now be seen
 as follows: making use of the identity~\cite{AJ}
 \footnote[2]{ The principal value prescription
 implicit in this identity is different from the one recently
 used in ref~\cite{Ram}.}

\begin{equation}
\int dx\phi (x) \Bigl( \pint dy { \phi (y) \over x-y } \Bigr)^2=
{\pi^2 \over 3} \int dx \phi^3(x) \label{Evui}
\end{equation}

 and using the symmetry of the solution, which implies that

\begin{equation}
\int dx{\phi_{0}\over x}=0, \label{Esci}
\end{equation}

 we obtain

\begin{eqnarray}
V_{eff}[\phi]&=&{1\over 2}\int dx\phi (x)\Bigr[ \pint dy {\phi (y)
\over
x-y}- \Bigl(\omega x- {\eta \over x} \Bigr) \Bigr]^2 \label{Ebcf}\\
\nonumber
&=& {\pi^{2}\over 6}\int dx\phi_{0}^{3}(x) -{\omega\over 2}
(\int dx\phi_{0}(x))^{2} \\
\nonumber
&&+{1\over 2}\int dx (\omega x-{\eta\over x})^{2}\phi(x)
\end{eqnarray}

 Thus the condition for the classical vacuum

\begin{equation}
\Bigl( {\delta V_{eff} \over \delta \phi}
\Bigr)_{\phi =\phi_0}=0 \label{Edov}
\end{equation}

 is equivalent to

\begin{equation}
\pi^{2}\phi_{0}^{2}=2\omega +2\omega\eta -\omega^2 x^2
-{\eta^2\over x^2}\label{Etcp}
\end{equation}

 (the right hand side now corresponds to a "$d=1$ potential")
 and reduces to a simple algebraic identity. This is a direct
 consequence of the symmetry of the classical vacuum. It is easy to
 verify that the solution to ~(\ref{Edov}) is in agreement with
 the analytic solution (40).

 We now give the collective field (40) an interpretation in terms of
 the density of zeroes of certain Laguerre polynomials. In terms of
 the original eigenvalue variables of the model, the equation for the
 vacuum~(\ref{Ecfv}) is:

\begin{equation}
\sum_{l=-{N\over 2} l\ne m}^{N\over 2}{1 \over x_m-x_l}
=\omega x_m -{\eta \over x_m} \label{Eito}
\end{equation}

 To solve this equation, we use the results of Calegero's
 work~\cite{Calo},
 based on the classical results of Stieltjies~\cite{Stie}. They
 show that the zeroes of the Laguerre polynomial
 $L_{N\over 2}^{\alpha}(x)$ obey:

\begin{equation}
\sum_{l=1 l\ne m}^{N\over 2} {1 \over x_m-x_l} ={1\over 2}
\Bigl( 1-{1 + \alpha \over
x_m} \Bigr) \label{Ecrs}
\end{equation}

 Using the symmetry of the solution to write

\begin{eqnarray}
\nonumber
\sum^{N\over 2}_{n=-{N\over 2} n\ne m}{1\over x_{m}-x_{n}}
&=& \sum^{N\over 2}_{n=1 n\ne m}\Bigl[ {1\over x_{m}-x_{n}}+
{1\over x_{m}+x_{n}}\Bigr] \\
&=& 2x_{m}\sum_{n=1 n\ne m}^{N\over 2} {1\over x^{2}_{m}-x^{2}_{n}}
\end{eqnarray}

 we find that the zeroes of $L^{\eta - 1}_{N\over 2} (\omega x^2)$
 obey~(\ref{Eito}). Thus we see that the collective field $\phi_{0}$
 in fact describes the density of zeroes of the Laguerre polynomials
 in the limit that these polynomials have an infinite number of
 nodes.
 In the case of the harmonic oscillator ($\eta =0$) Jevicki and
 Levine~\cite{JeLe} proved explicitly that the
 collective field describes the density of zeroes of the Hermite
 polynomials which are of course special cases of the Laguerre
 polynomials.

\section{The Spectrum}

 In this section we study the spectrum of the model. When
 supersymmetry
 is not broken in the leading order, we can expand as
 in~(\ref{Eevc}) and
 obtain the following bosonic and fermionic quadratic contributions
 to the Hamiltonian

\begin{eqnarray}
\nonumber
H_0^B &=&{1\over 2}\int dx\phi_0 p^2 -{1\over 2}\pint dxdy
{v(x)-v(y) \over x-y}
 \partial_x\eta\partial_y\eta \\
&+&{\pi^2\over 2} \int dx\phi_0 (\partial_x\eta)^2 \\
\nonumber
H_0^F &=&{1\over 2}\int dx\Bigl[\psi^\dagger (x)
\sqrt{\phi_0 (x)},{d\over dx}
\pint dy{\psi(y)\sqrt{\phi_0 (y)}\over x-y} \Bigr] \label{Ebqh}
\end{eqnarray}

 Considering the bosonic sector of the quadratic Hamiltonian, we
 see that
 apart from the term

\begin{equation}
-{1\over 2}\pint dxdy {v(x)-v(y) \over x-y}\partial_x
\eta \partial_y \eta \label{Eets}
\end{equation}

 we have just got the terms associated in the bosonic string theory
 to a
 massless scalar particle~\cite{DaJ}. This term could, in principle,
 modify the spectrum in an unexpected way. For the potential we
 consider
 here, using the fact that $\int dx\partial_x\eta=0$,this term may be
 simplified to:

\begin{equation}
-{1\over 2} \Bigl[\int dx{\partial_x\eta\over x} \Bigr]^2
\label{Etds}
\end{equation}

 This term can be made to vanish if we systematically restrict
 ourselves
 to odd fluctuations $\eta$, which we assume from now on. Notice
 that
 from~(\ref{Eevc}) this simply implies that our density field
 $\phi (x)$
 is always even, which is natural in view of the ground state of the
 previous section. Thus, under this restriction, the
 bosonic spectrum is
 that of a massless scalar. In order to exhibit this explicitly, one
 changes to the time of flight variable $q$~\cite{DaJ}

\begin{equation}
dq = {dx \over \phi_0} {\hskip.5in} q=\int
{dx \over \phi_0(x)} \label{Ectc}
\end{equation}

 and rescales

\begin{equation}
p \to {p \over \phi_0} \hskip.5in \psi \to
{\psi \over \sqrt\phi_0} \label{Ertv}
\end{equation}

 The change of co-ordinates~(\ref{Ectc}) is easily integrated to
 yield an
 explicit expression for $q$:

\begin{equation}
q = {\pi \over 2\omega} arcos \Bigl[ {-\omega\over\sqrt{1+2\eta}}
\Bigl( x^2-{1+\eta\over\omega}\Bigr)\Bigr] \label{Eeeq}
\end{equation}

 where in the above, we have chosen $q=0$ to correspond to
 $x^{2}=y_{-}^{2}$.

 A few remarks about the new parametrization~(\ref{Ectc}) are in
 order. First, recall that in $x$ space $\phi_{0}$ is defined on two
 disconnected intervals. Since we have picked $q=0$ to correspond to
 $x^{2}=y_{-}^{2}$, $\phi_{0}$ is defined on a single interval in $q$
 space. In terms of the new co-ordinate $q$, integration from $y_{-}$
 to $y_{+}$ corresponds to integrating from $q=0$ to
 $q=L={\pi^{2}\over 2\omega}$, and integration from $-y_{+}$ to
 $-y_{-}$ corresponds to integrating from $q=-L$ to $q=0$. Explicitly

\begin{eqnarray}
\nonumber
x(q)=\epsilon (q)\Bigl[ {1+\eta \over \omega}-{\sqrt{1+2\eta}
\over\omega}
 \cos ({2\omega q\over\pi})\Bigr]^{1\over 2},\\
 -L\le q\le L \label{Cord}
\end{eqnarray}

 In $q$-space, for odd fluctuations $\eta$, we obtain:

\begin{equation}
H_0^B={1\over 2}\int dq\Bigl( p^2+\pi^2(\partial_q\eta)^2\Bigr)
\label{Efqb}
\end{equation}

 If we now expand our fields in terms of the oscillators

\begin{equation}
\eta (q) = \sum_{n=1}^{\infty} {1\over\sqrt{4\pi^2 n}}(a_n +
a_n^\dagger )
   sin{n\pi q\over L} \label{Eetf}
\end{equation}

\begin{equation}
p(q)=\sum^{\infty}_{n=1}-i\sqrt{\pi^2 n\over 4L^2}(a_n -a_n^\dagger )
 sin{n\pi q\over L} \label{Etep}
\end{equation}

 where $[a_{m},a_{n}]=\delta_{mn}$, we obtain

\begin{equation}
H_0^B=\sum_{n=1}^{\infty} n\omega (a_n^\dagger a_n+{1\over 2})
\label{Efos}
\end{equation}

 The fermionic sector of the quadratic Hamiltonian in $q$-space is

\begin{equation}
H_0^F={1\over 2}\int^{L}_{-L} dq\Bigl[\psi^\dagger (q),{d \over dq}
\pint^{L}_{-L} dq'{\phi_0(q')\psi (q') \over x(q)-x(q')}\Bigr]
\label{Eswf}
\end{equation}

  The identity:

\begin{equation}
\pint^{L}_{-L} dq'{\phi_0(q')\psi (q') \over x(q)-x(q')} =
\pint^{L}_{-L} dq'{\phi_{0}(q')x(q')\psi (q')\over x^{2}(q)-
x^{2}(q')}
\label{ident}
\end{equation}

 holds for our potential $v(x)$, provided we restrict ourselves to
 odd
 fermionic fluctuations $\psi$. On the left hand side of the above
 equation, there is a singularity of the form $\sqrt{1 \over q}$
 at $q=-{i\pi \over 2\omega}log {1\over\sqrt{1+2\eta} }$
 arising form the $1\over x$ factor in $\phi_{0}(x)$.
 This singularity does not contribute for odd fermionic fluctuations
 $\psi $ which is evident since, the factor $x(q')$ cancels this
 singularity on the right hand side of the above equation. Thus we
 again see
 the importance of the symmetry of the solution.

 To determine the fermionic sector of the spectrum, introduce the
 oscillators

\begin{eqnarray}
\nonumber
\psi (q)&=&{1\over\sqrt{2L}}\sum_{n=1}^{\infty} b_n sin{n\pi q\over
L} \\
\nonumber
\psi^\dagger (q) &=&{1\over\sqrt{2L}}\sum_{n=1}^{\infty} b_n^\dagger
  sin{n\pi q\over L} \label{Etfe}
\end{eqnarray}

 where $\{b_m,b^\dagger_n \} = \delta_{mn}$.
 From~(\ref{ident}), we find that we need to compute integrals of the
 form:

\begin{equation}
\pint^{L}_{-L}dq' {\phi_{0}(q')x(q')\sin ({n\pi q'\over L})\over
x^{2}(q)-x^{2}(q')}  \label{inttocomp}
\end{equation}

 In the above replace $\sin({n\pi q'\over L})$ by
 ${1\over i}e^{in\pi q'\over L}$. This integral is now easily
 computed using contour integration arguments.
 Exploit the fact that the integrand is periodic
 with period $2L$, by closing the contour in the upper half
 plane, along the lines $q=-L$ and $q=L$.
 Close the box at infinity, parallel to the $q$ axis.
 The only poles that contribute are on the real line at
 $q$ and $-q$. This is only true for restricted
 fluctuations, since in general the $1\over x$ singularity
 would contribute. We employ a principal value prescription
 which corresponds to taking one half of the residues of poles
 lying on the contour. The integrals are now easily computed, which
 together with the bosonic contribution gives the quadratic
 Hamiltonian:

\begin{equation}
H_0 = \sum_{n=1}^{\infty}
n\omega (a_n^\dagger a_n + b_n^\dagger b_n) \label{Efsq}
\end{equation}

 thus explicitly demonstrating the supersymmetry of the semiclassical
 spectrum.

 The above discussion has been carried out in $q$ space, or time of
 flight
 variable. Following~\cite{JeRo} the $x$ space kernel that
 corresponds to the
 second of (49), can be written as

\begin{equation}
\int dx \pint dy { [\psi^\dagger (x),\psi(y)] \over (x-y)^2 } \phi_0
(y) \label{Errw}
\end{equation}

 Thus we see that by studying the fermionic fluctuations,we are
 naturally
 led to study the operator:

\begin{equation}
(O\psi )(x) =\pint dy{\psi(y)\phi_0 (y)\over (x-y)^2} \label{Eots}
\end{equation}

 The above integral runs over the two disconnected bands of
 eigenvalues
 described by $\phi_0 (x)$. The eigenfunctions are most easily
 determined
 by performing this integral in $q$ space:

\begin{eqnarray}
\nonumber
\pint dy {\psi(y)\phi_{0}(y)\over (x-y)^{2}}&=&{d\over dx(q)}
\pint^{L}_{-L}dq'{\phi^{2}_{0}(q')\psi(q')\over x(q')-x(q)} \\
&&={1\over\phi_{0}(q)}{d\over dq}
\pint^{L}_{-L}dq'{\phi^{2}_{0}(q')\psi(q')\over x(q')-x(q)}
\end{eqnarray}

 Making the ansatz $\psi (q)\phi_{0} (q)=\sin {n\pi q\over L}$, this
 integral reduces to~(\ref{inttocomp}). Explicitly performing the
 integral,
 we find

\begin{equation}
\psi_{n}(q)={\sin ({n\pi q\over L})\over\phi_{0}(q)}
\end{equation}

 is an eigenfunction with eigenvalue $n\omega$. Rewriting this in
 $x$ space, we find

\begin{equation}
\psi_{n}(x)={\pi x\over \sqrt{1+2\eta}}B_{n-1}(
{1+\eta\over\sqrt{1+2\eta}}-{\omega x^{2}\over\sqrt{1+2\eta}})
\end{equation}

 with $B_{n}$ a Chebyshev polynomial. Notice that these
 eigenfunctions
 are indeed odd on the two bands. This is not suprising, since in the
 case of the harmonic oscillator ($\eta = 0$), the eigenfunctions are
 simply the Chebyshev polynomials~\cite{JeRo}.

 A class of integral operators of the type~(\ref{Eots}) has been
 studied
 by Calogero~\cite{Calo}. To make a connection with his work,
 rescale
 $x^{2}\to{\sqrt{1+2\eta}\over\omega}x^{2}$. The
 kernel~(\ref{Eots})
 becomes, after some algebra

\begin{equation}
(O\psi )(x)={d\over dx}{\pint^{a_{+}}}_{a_{-}}dy^{2}{
\sqrt{1-(y^{2}-{1+\eta\over\sqrt{1+2\eta}})^{2}}\psi(y)\over
\pi y(x^{2}-y^{2})}
\end{equation}

 with $a_{\pm}=(1+2\eta)({1+\eta\over\sqrt{1+2\eta}}\pm 1)$. The
 above analysis shows that the eigenfunctions $\psi_{n}$

\begin{equation}
\psi_{n}(x)=xB_{n-1}({1+\eta\over\sqrt{1+2\eta}}-x^{2})
\end{equation}

 of the kernel have integer eigenvalues. Explicitly, the first
 three eigenfunctions are:

\begin{eqnarray}
\nonumber
\psi_{1}(x)&=&x\\
\nonumber
\psi_{2}(x)&=&x({1+\eta\over\sqrt{1+2\eta}} -x^{2})\\
\nonumber
\psi_{3}(x)&=&x(1+2({1+\eta\over\sqrt{1+2\eta}} -x^{2})^{2})
\end{eqnarray}

 This generalizes well known results involving kernels of Chebyshev
 polynomials with weight $\sqrt{1-x^{2}}$~\cite{AbSt}.

\section{Interactions}

 We have seen that by suitably restricting
 the fluctuations, one is able to avoid a singularity at the origin
 in
 $q$ space, at the level of the quadratic Hamiltonian.
 We were able to show that supersymmetry is unbroken, by computing
 the
 spectrum. This must indeed be the case, since in~\cite{RoVa},
 the authors
 argued that if a supersymmetric classical configuration can be
 found,
 then the semiclassical spectrum is supersymmetric. Under this
 restriction
 on the fluctuations, it turns out that
 this singularity does not contribute in {\it any}
 higher order interactions~\cite{RoDe}. In this section,
 we illustrate this for the cubic interaction.

 The cubic contribution to the Hamiltonian is:

\begin{eqnarray}
H_{3}&=&{1\over 2N}\int dx (\partial_{x}\eta )p^{2} \\
\nonumber
&&+{1\over 2N}\int dx \partial_{x}\eta \Bigl(\pint dy{\partial_{y}
\eta\over
x-y}\Bigr)^{2} \\
\nonumber
&&-{i\over 2N}\int dx[(\partial_{x}\psi^{\dagger})\psi-\psi^{\dagger}
(\partial_{x}\psi )]p \\
\nonumber
&&-{1\over 2N}\int dx [\psi^{\dagger},\psi ]{d\over dx}\pint dy
{\partial_{y}\eta\over x-y}
\end{eqnarray}

  In $q$ space, this becomes, after some algebra:

\begin{eqnarray}
H_{3}&=&{1\over 2N}\int^{L}_{-L} dq (\partial_{q}\eta )p^{2} \\
\nonumber
&&+{1\over 2N}\int^{L}_{-L} dq \partial_{q}\eta
\Bigl(\pint^{L}_{0} dq'{2x(q)\partial_{q'}\eta (q')\over x^{2}(q')
-x^{2}(q)}\Bigr)^{2} \\
\nonumber
&&-{i\over 2N}\int {dq\over \phi_{0}^{2} (q)}
[(\partial_{q}\psi^{\dagger})\psi-\psi^{\dagger}
(\partial_{q}\psi )]p \\
\nonumber
&&+{1\over 2N}\int^{L}_{-L} dq [\psi^{\dagger},\psi ]
\pint^{L}_{-L} dq'
{\partial_{q'}\eta\over (x(q)-x(q'))^{2}}
\end{eqnarray}

 Notice that in the above there are no factors of $\phi_{0}(q)$. Thus
 the $\sqrt{1\over q}$ singularity does not appear. The only new
 poles
 are introduced by the $1\over\phi^{2}_{0}$ factor appearing in the
 last term. These new poles all lie on the real line, at
 $q=0,\pm L$ and the integrals are easily computed. The principal
 valued
 integrals are performed by closing as for~(\ref{inttocomp}). Our
 principal value prescription corresponds to taking one quarter of
 the
 residues of the poles lying on the corners ($q=\pm L$) of the
 contour.
 The cubic interaction in the oscillator basis is now easily found:

\begin{eqnarray}
\nonumber
H_{3}&=&{1\over 4N\sqrt{\pi L}} ({1+\eta \over 1+2\eta})
\sum_{n,p,y,l>0}'
\sqrt{k_{l}k_{n}k_{p}}(|k_{l}+k_{n}+k_{p}|\\
&&-|k_{l}-k_{n}-k_{p}|)a_{n}^{\dagger}a_{p}^{\dagger}a_{l}+\\
\nonumber
&&+{1\over 8N\sqrt{\pi L}} ({1+\eta \over 1+2\eta})\sum_{n,p,y,l>0}'
\sqrt{k_{l}}(k_{p}-k_{n})\\
\nonumber
&&(|k_{l}+k_{n}+k_{p}|-|k_{l}+k_{n}-k_{p}| +|k_{l}-k_{n}+k_{p}|\\
\nonumber
&&-|k_{l}-k_{n}-k_{p}|)b_{n}^{\dagger}b_{p}a_{l}^{\dagger}
+h.c.
\end{eqnarray}

 where the prime on the sum indicates a sum over even integers only,
 and
 $h.c.$ denotes the hermitean conjugate.

\section{Discussion and conclusions}

We have carried out a study of the supersymmetric collective
field theory in the case of a harmonic potential deformed by
a singular term.
The emphasis has been on the density description of the underlying
super Calogero model. The restriction on the density
consistent with supersymmetry has been identified and forces one to
consider systematically two band ansatz. Several classical results
of Calogero both at the level of leading classical configuration
and fluctuations have been obtained.

In the context of Penner models Tan~\cite{Tan} identified a double
scaling limit in terms of an analytically continued crossover of the
two
bands. This analytic continuation cannot be performed in our
model. Therefore the existence of a double scaling limit in
Marinari-Parisi type models that does not simply amount to
stabilization of
$d=0$ models remains an open question.


\end{document}